\documentstyle[prc,aps,preprint]{revtex}

\begin{document}
\draft
\tighten

\title{Equilibrium Free Energies from Nonequilibrium 
Processes\footnote{
Talk presented during the
Marian Smoluchowski Symposium on Statistical Physics
in Zakopane, Poland, September 1-10, 1997.}}
\author{C. Jarzynski}
\address{Theoretical Astrophysics, T-6, MS B288 \\
         Los Alamos National Laboratory \\
         Los Alamos, NM 87545 \\
         {\tt chrisj@t6-serv.lanl.gov}}
\date{\today}

\maketitle

\begin{abstract}
A recent result, relating the (irreversible)
work performed on a system during a 
non-quasistatic process, to the Helmholtz free 
energy difference between two equilibrium states 
of the system, is discussed.
A proof of this result is given for the special 
case when the evolution of the system in 
question is modelled by a Langevin
equation in configuration space.
\end{abstract}

My purpose in this talk is to present and
discuss a result which relates the free energy
difference between two equilibrium states of
a system --- defined with respect to two
values of an external parameter ---
to the work performed on the system by changing 
that parameter {\it at a finite rate} from one
value to the other.

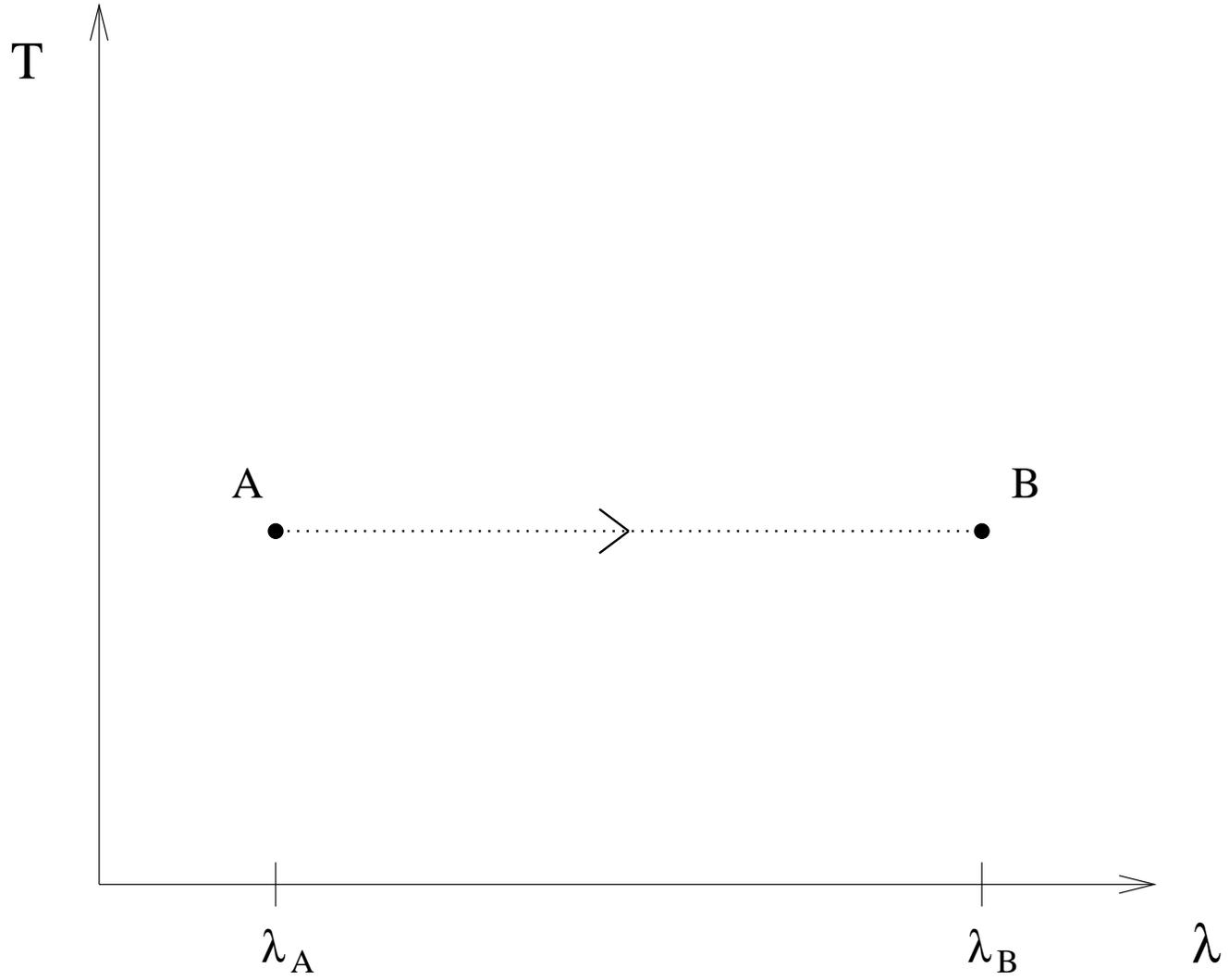
\begin{figure}
\caption{
The unique macroscopic equilibrium state
corresponding to a temperature $T$,
and a value of the external
parameter $\lambda$, may be represented by
a point in the $(\lambda,T)$-plane.
By keeping the system thermostatted 
at a constant temperature, and changing the 
external parameter quasistatically,
we can reversibly switch the system from one
equilibrium state, $A$, to another, $B$.}
\label{fig1}
\end{figure}

Let me begin with a fundamental
statement from classical thermodynamics\cite{ll}:
the total work performed on a system during
an isothermal, {\it quasistatic} process 
is equal to the free energy difference
between the initial and final equilibrium
states of the system.\footnote{
Throughout this talk, the term ``free
energy'' will refer specifically to the
{\it Helmoltz} free energy.}
This statement may be understood as follows.
Imagine a finite system which depends on
some external parameter, $\lambda$.
Macroscopically, an {\it equilibrium state} 
is the unique state attained by the system
by allowing it to come to 
equilibrium with an infinite heat 
reservoir at temperature $T$, holding
$\lambda$ fixed.
Such a state may be
represented by a single point in the
$(\lambda,T)$-plane, as shown in 
Fig.\ref{fig1}.
With each equilibrium state we may associate
a free energy $F$:
\begin{equation}
F(\lambda,T) = E-ST,
\end{equation}
where $E$ and $S$ denote, respectively, 
the internal energy and the entropy of
the system, both functions of the state.
If we now prepare the system in a state
$(\lambda_A,T)$, and then infinitely
slowly change the value of $\lambda$ from
$\lambda_A$ to $\lambda_B$, always keeping
the system thermostatted (i.e.\ in contact
with the heat reservoir) at temperature $T$,
then the system will 
evolve through a continuous sequence of
equilibrium states --- represented by
the dotted line in Fig.\ref{fig1} --- and the
net external work which we perform 
during this process will equal
the free energy difference between the
initial and final states:
\begin{equation}
\label{eq:qstat}
W_\infty = \Delta F \equiv
F(\lambda_B,T) - F(\lambda_A,T).
\end{equation}
The subscript on $W$ reminds us that this 
process is carried out quasistatically.

At the microscopic level, we must treat
the system statistically, replacing 
the unique macroscopic state by a 
probability distribution (or ensemble)
of micro-states of the system.
An equilibrium state of the sort discussed 
in the previous paragraph is 
represented by a canonical ensemble in
the microscopic phase space, and the
free energy is given by
\begin{equation}
F(\lambda,T) = -k_B T\ln Z(\lambda,T),
\end{equation}
where $Z$ denotes the partition function.
During a process in which the system remains 
thermostatted while $\lambda$ is varied 
quasistatically from $\lambda_A$ to $\lambda_B$, 
the statistical state of the system evolves
through a continuous sequence of canonical
ensembles, and again the work performed is
equal to the free energy difference between
initial and final equilibrium states:
$W_\infty = \Delta F$.

Eq.\ref{eq:qstat} is thus a basic statement 
from both macroscopic thermodynamics and 
microscopic statistical mechanics.
The central point which I wish to make
is that there exists a comparably simple
result, which relates $\Delta F$ ({\it defined}
as the free energy difference between two
equilibrium states $A$ and $B$) to the
work performed during a process in which
$\lambda$ is changed from $\lambda_A$ to
$\lambda_B$ at a {\it finite} rate; hence, 
a {\it nonequilibrium}, irreversible process.

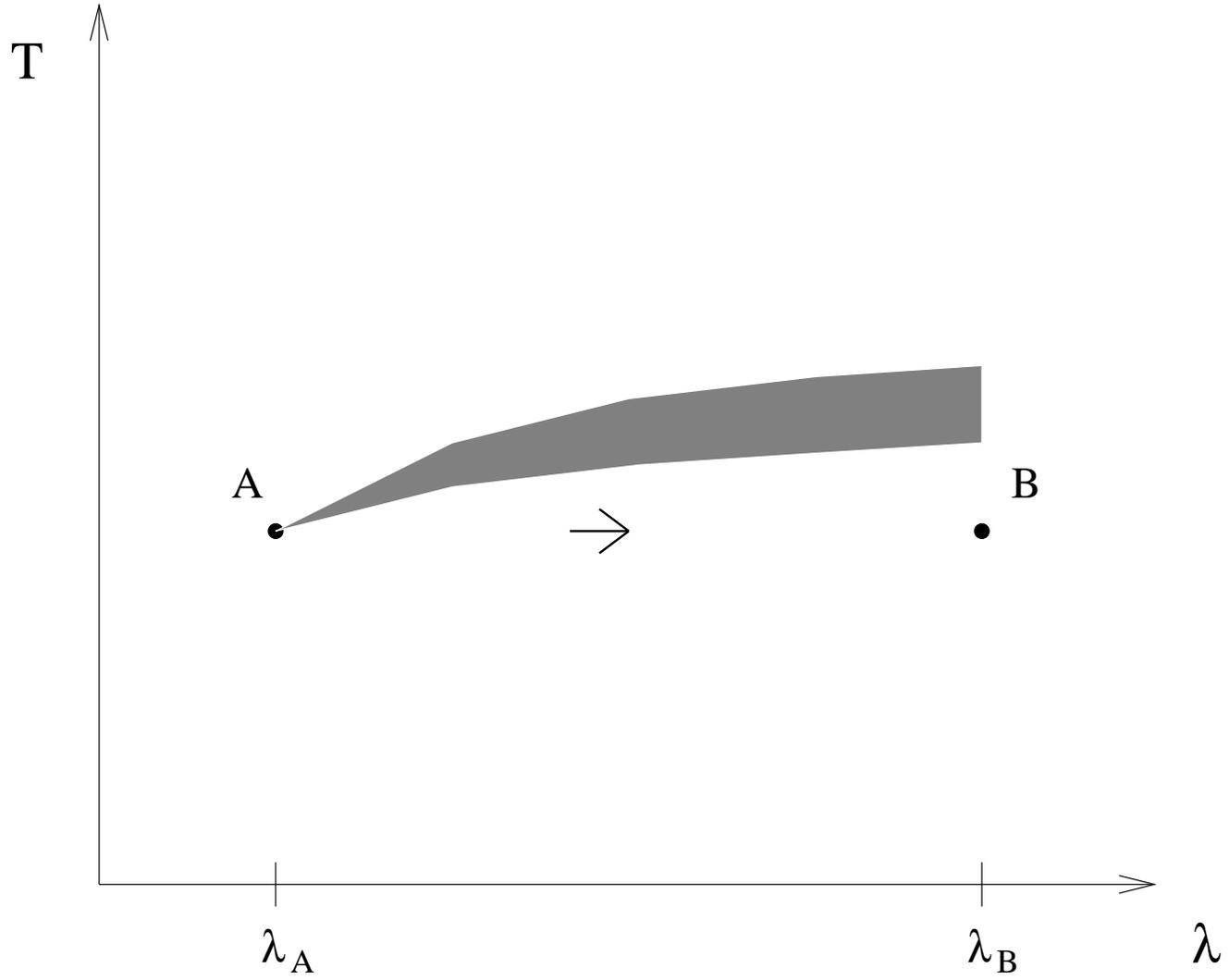
\begin{figure}
\caption{
If we start with the system in equilibrium 
state $A$, but then switch the external parameter
from $\lambda_A$ to $\lambda_B$ at a {\it finite}
rate, the system will progress through
a sequence of {\it nonequilibrium} states, 
indicated schematically by the shaded region.}
\label{fig2}
\end{figure}

Consider therefore the following sequence of steps.
(1) With $\lambda$ fixed at an initial value
($\lambda_A$), we let the system equilibrate 
with a reservoir at temperature $T$.
(2) We then externally ``switch''
$\lambda$ from the initial
value ($\lambda_A$) to a final one 
($\lambda_B$) over a finite time $\tau$.
(For specificity, assume 
$\dot\lambda\equiv d\lambda/dt$
to be constant.)
The system remains in contact
with the reservoir as we switch $\lambda$.
(3) Once $\lambda$ reaches its destination
($\lambda_B$), we note down the external 
work $W$ which we performed on the system 
during this process.
(4) Go back to step 1 and repeat {\it ad
infinitum}.
Steps 1-3 define what I will call a single
{\it realization} of the switching process;
by virtue of step 4, we obtain a 
{\it statistical ensemble} of such 
realizations.
Note that although the system begins in
equilibrium with the reservoir, it does not
generally remain so once we start changing
$\lambda$ at a finite rate.
Typically, the system will be found in a 
{\it nonequilibrium} state, not characterized
by a single point $(\lambda,T)$
(though the value of $\lambda$ is of course
well defined at every instant).
Schematically, I have depicted this situation
by representing the evolution of the system
as a shaded area, see Fig.\ref{fig2}, rather 
than as a single line as in Fig.\ref{fig1}.

(At the end of the switching process, we can
of course always opt to hold the value of
$\lambda$ fixed, at $\lambda_B$, and allow
the system to relax to equilibrium with the
reservoir, in this way finally attaining the
equilibrium state $B$. 
This then gives us a prescription for going
from one equilibrium state, $A$, to another,
$B$, via a nonequilibrium process.
However, once we stop changing the external
parameter $\lambda$, we stop performing 
external work on the system.
Therefore the central result to be presented
below, which makes a statement about the work
$W$ performed externally, is independent of
whether or not we carry out the 
supplementary step of allowing the system to
relax to state $B$ at the end of the
switching process.)

Since the switching described in step 2 above
is carried out at a finite rate, we expect
statistical fluctuations: the work $W$ will
differ from one realization to the next.
Thus, having obtained a statistical ensemble
of realizations by repeating steps 1-3
{\it ad nauseum}, we will have a distribution
of values of work, $\rho(W)$, defined such
that $\rho(W)dW$ gives the proportion of
realizations for which the work fell within
an infinitesimal window $dW$ around a 
particular value $W$.

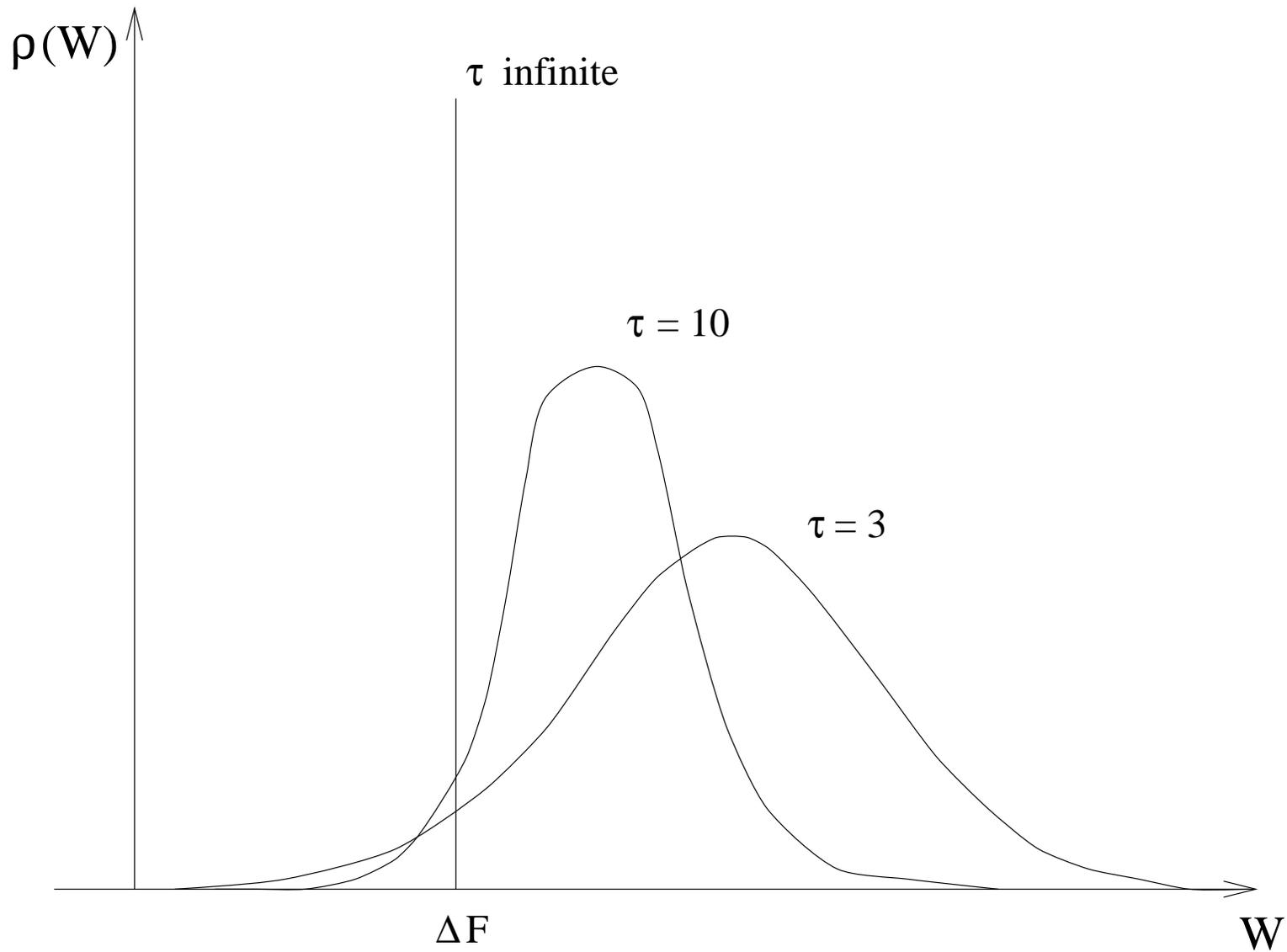
\begin{figure}
\caption{
The distribution of values of work, $\rho(W)$, 
performed during a statistical ensemble of 
switching processes, depends on the switching 
time $\tau$.
For $\tau\rightarrow\infty$ this distribution
becomes a $\delta$-function at
$W=\Delta F$.
Generically, one expects a broader distribution,
the more rapidly the switching is performed.}
\label{fig3}
\end{figure}

The distribution $\rho(W)$ will depend on how 
slowly or quickly the switching was performed, 
that is, on the switching time $\tau$.
(In carrying out steps 1-4 above, it was
assumed that $\tau$ remained the same 
from one realization to the next.)
For $\tau\rightarrow\infty$, we get
$\rho(W)\rightarrow\delta(W-\Delta F)$,
by Eq.\ref{eq:qstat},
whereas for $\tau$ finite we expect the
distribution $\rho(W)$ to have a finite width.
Generically, the faster we switch $\lambda$,
the larger the expected fluctuations in $W$,
as depicted in Fig.\ref{fig3}.

One final point before I reach the punch line.
While $W=\Delta F$ in the quasistatic, 
reversible limit ($\tau\rightarrow\infty$),
for {\it finite} switching times we expect
the work performed to exceed the free energy
difference $\Delta F$:
\begin{equation}
\label{eq:ineq}
\overline{W} \ge \Delta F,
\end{equation}
where the overbar denotes an average over our
statistical ensemble of realizations of the
switching process, carried out at a fixed
value of $\tau$.
(This statement does not preclude an
occasional measurement of $W$ which falls
{\it below} $\Delta F$, though for macroscopic
systems such fluctuations will be 
exceedingly rare.)
The inequality given by Eq.\ref{eq:ineq}
essentially follows from the Second Law:
the work performed during an irreversible
process is expected to exceed
that performed during the corresponding
reversible process.

Let me now make a claim which is the
central focus of this talk.
If instead of taking the average of the 
work $W$, over a statistical ensemble of
switching realizations,
we take the average of the quantity
$e^{-\beta W}$, where $\beta\equiv 1/k_BT$,
then {\it that average will be equal to}
$e^{-\beta\Delta F}$, for {\it any} value
of the switching time $\tau$:
\begin{equation}
\label{eq:iden}
\overline{\exp -\beta W}\equiv
\int dW\,\rho(W,\tau)\,\exp -\beta W 
= \exp -\beta\Delta F
\qquad,\qquad {\rm for\, all}\,\, \tau.
\end{equation}
(The dependence of $\rho(W)$ on the switching
time $\tau$ has been made explicit here.)
That is, the average of $e^{-\beta W}$
will be the same over, say, any of the three
distributions shown in Fig.\ref{fig3}, and will equal
$e^{-\beta\Delta F}$.

Eq.\ref{eq:qstat} relates the free energy 
difference between two equilibrium states of a 
system, to the work performed in {\it reversibly}
taking the system from one state to the other.
Eq.\ref{eq:iden} is the extension of this
statement to {\it irreversible} 
(nonequilibrium) processes.
Note that, since $\Delta F$ depends only on 
the equilibrium states $A$ and $B$, Eq.\ref{eq:qstat}
implies that the reversible work performed in
going from $A$ to $B$ is independent of the path
taken from $\lambda_A$ to $\lambda_B$ in 
parameter space.
(We have imagined only a single parameter
$\lambda$, but of course more generally
parameter space can be multi-dimensional.)
Eq.\ref{eq:iden} makes a comparable statement as 
regards the {\it rate} at which we switch $\lambda$:
it says that the ensemble average of 
$e^{-\beta W}$ is independent, not only of
the path from $\lambda_A$ to $\lambda_B$, 
but also of how quickly or slowly we vary
$\lambda$ along that path.

Eq.\ref{eq:iden} gives the relationship between a 
quantity defined with respect to {\it equilibrium}
states of a system ($\Delta F$) to
a quantity extracted from an ensemble of
{\it nonequilibrium} processes.
Moreover, this relationship takes the form of
an {\it equality}, whereas
most statements relating equilibrium and
nonequilibrium quantities are expressed as
inequalities, for instance, Eq.\ref{eq:ineq}.
Indeed, as I have pointed out elsewhere\cite{iden}, 
the inequality $\overline{W}\ge\Delta F$ follows
immediately and rigorously from Eq.\ref{eq:iden}.

As a quick consistency check, we can verify the
validity of Eq.\ref{eq:iden} in two limiting cases:
infinitely slow ($t_{s}\rightarrow\infty$) and
infinitely fast ($t_{s}\rightarrow 0$) switching.
In the former case, we get 
$\rho\rightarrow\delta(W-\Delta F)$,
and Eq.\ref{eq:iden} is satisfied.
For $t_{s}\rightarrow 0$, as discussed 
elsewhere\cite{iden},
Eq.\ref{eq:iden} reduces to the following
well-known identity:
\begin{equation}
\langle\exp -\beta\Delta H\rangle_{A}=
\exp -\beta\Delta F,
\end{equation}
where $\Delta H=H_{B}-H_{A}$ is the difference
between the initial and final Hamiltonians,
and $\langle\cdots\rangle_{A}$ defines a
canonical average over the initial equilibrium
state.

A proof of Eq.\ref{eq:iden}, based on a treatment
in which both the system of interest and the heat
reservoir are explicitly taken into consideration,
was given in Ref.\cite{iden}.
Another proof, based on a master equation approach, 
was given in Ref.\cite{mast}, along with supporting 
numerical results.
Yet a third proof, assuming Markovian evolution
and microscopic reversibility (and which along the
way yields an interesting detailed balance relationship
for multiple time-step processes), has been found
by Gavin Crooks\cite{gavin}.

In the spirit of this Symposium, I will 
sketch a proof of Eq.\ref{eq:iden} for
the case in which the evolution of the system of
interest is modelled by a Langevin equation in
configuration space,
thus a statistical ensemble of such systems 
evolves under a Smoluchowski equation.
(Physically, this corresponds to the limit of
overdamped evolution, in which the momentum of
the system reaches equilibrium with the heat
reservoir on a time scale very short compared
both to that required for the configuration of the
system to equilibrate, and to the time $\tau$
over which we perform the switching.)
Although this case is a particular example of the 
situation considered in Ref.\cite{mast}, 
the proof presented here is different 
from those of Refs.\cite{iden,mast,gavin}.

Let us assume that the Hamiltonian for our system
has the form
$H_\lambda(x,p) = p^2/2m + V_{\lambda}(x)$.
(Although we assume a one-dimensional
configuration space, the generalization 
to more degrees of freedom is straightforward.)
The free energy difference may then be
expressed in terms of a ratio of configurational
partition functions:
\begin{eqnarray}
\Delta F &=& -\beta^{-1}\ln
{Q_{B}\over Q_{A}} \\
Q_{\lambda}&=&\int dx\,
\exp -\beta V_{\lambda}(x).
\end{eqnarray}
(I will often use $A$ and $B$
in place of $\lambda_A$ and $\lambda_B$, respectively.)
Let us now take the evolution of the configuration
of the system to obey a Langevin equation:
\begin{equation}
\label{eq:lang}
\dot x = v_{\lambda}(x) +
\tilde u(t),
\end{equation}
where $\tilde u(t)$ is a term representing white
noise,
\begin{equation}
\langle \tilde u(t)\tilde u(t+s)\rangle =
D\delta(s),
\end{equation}
with $\langle\cdots\rangle$ denoting an average over
realizations of the noise;
and $v_\lambda(x)$ is the terminal velocity
attained by a particle subject to both a
conservative force $-\partial V_\lambda/\partial x$,
and a frictional force $-\gamma\dot x$
satisfying the Einstein relation,
$\gamma^{-1} = \beta D/2$:
\begin{equation}
v_{\lambda}(x) = -{\beta D\over 2}
{\partial V_{\lambda}\over\partial x}
(x).
\end{equation}
In Eq.\ref{eq:lang}, the external parameter
depends on time, according to
\begin{equation}
\label{eq:loft}
\lambda(t) = \lambda_{A} +
(\lambda_{B}-\lambda_{A}) t/\tau.
\end{equation}

A stochastic trajectory $x(t)$, $t\in [0,\tau]$, 
satisfying Eq.\ref{eq:lang}, represents the evolution 
of our system during a single realization of the
switching process.
We will now make use of the fact that, for
such evolution, one can write down an explicit
expression for the probability distribution of
trajectories $x(t)$.
Namely, given an initial condition $x(0)=x_{0}$,
the probability for obtaining a particular
trajectory $x(t)$, as we switch $\lambda$ from
$A$ to $B$, is given by:\cite{polymers}
\begin{eqnarray}
\label{eq:pab}
P_{A\rightarrow B}[x(t)] &=& {\cal N}
\exp -S_{+}[x(t)]\\
\label{eq:sdef}
S_{\pm}[x(t)] &=& {1\over 2D}\int_{0}^{\tau} dt\,
\Bigl(\dot x \pm {\beta D\over 2}
\partial_{x} V_{\lambda}\Bigr)^2,
\end{eqnarray}
where $\lambda=\lambda(t)$ in the integrand
of Eq.\ref{eq:sdef}.
(I will make use of $S_{-}$ shortly.)
The normalization factor ${\cal N}$ is chosen so
that the integral of $P_{A\rightarrow B}$
over all trajectories $x(t)$,
$t\in[0,\tau]$, starting from $x(0)=x_{0}$,
is unity:
\begin{equation}
\label{eq:normal}
\int{\cal D}_0[x(t)]\,P_{A\rightarrow B}[x(t)]=1.
\end{equation}
[The subscript on ${\cal D}$ indicates that we
are integrating over paths $x(t)$ with a fixed
initial point, $x(0)=x_0$.]

To make sense of Eqs.\ref{eq:pab} to \ref{eq:normal},
we must introduce a measure on path space.
We do this by dividing the interval $[0,\tau]$
into sub-intervals of duration
$\delta t=\tau/N$, then replacing the
continuous trajectory $x(t)$ by the discrete
set of configurations $x_n=x(t_n)$ at times
$t_n=n\delta t$ ($n=0,1,\cdots,N$), and at
the end taking the limit $N\rightarrow\infty$.
A convenient representation of the integral
in Eq.\ref{eq:sdef} is then:
\begin{equation}
S_{\pm}[x(t)] \rightarrow {1\over 2D}
\sum_{n=1}^N \delta t \,
\Bigl[
{x_n-x_{n-1}\over\delta t} \pm
{\beta D\over 2} \partial_x V_\lambda
(x_\nu)\Bigr]^2 
\qquad,\qquad \nu = n-{1\over 2}\mp{1\over 2}.
\end{equation}
In Eq.\ref{eq:normal}, the integral in path 
space, over all trajectories $x(t)$
with a fixed initial point $x(0)=x_0$, is
expressed as:
\begin{equation}
\int{\cal D}_0[x(t)] \cdots = 
\prod_{n=1}^N\int dx_n \cdots.
\end{equation}
It is straightforward to verify that the
normalization constant in Eq.\ref{eq:pab}
is given by ${\cal N}=(2\pi D\delta t)^{-N/2}$, 
in this scheme.
For future use, let us also define
$\int{\cal D}_\tau[x(t)] = 
\prod_{n=0}^{N-1}\int dx_n$, 
representing an integral over trajectories
$x(t)$ with a common {\it final} point
$x_\tau\equiv x(\tau) = x_N$.
Note that 
\begin{equation}
\label{eq:equivints}
\int dx_0\int {\cal D}_0[x(t)] \cdots =
\int dx_\tau\int {\cal D}_\tau[x(t)]\cdots =
\prod_{n=0}^N\int dx_n \cdots.
\end{equation}

For a particular trajectory $x(t)$, the external 
work performed on the system is:
\begin{equation}
\label{eq:work}
W[x(t)] = \int_{0}^{\tau} dt\,
\dot\lambda \,\partial_\lambda V_\lambda
\Bigl(x(t)\Bigr).
\end{equation}
If we now launch an ensemble of such trajectories, from
a canonical ensemble of initial conditions (corresponding 
to $\lambda=A$) at $t=0$, then an explicit
expression for the average of $e^{-\beta W}$ over
this statistical ensemble is given by:
\begin{equation}
\label{eq:explicit}
\overline{\exp -\beta W} =
\int dx_{0} {1\over Q_{A}}
\exp -\beta V_{A}(x_{0})\,
\int {\cal D}_0[x(t)]\,{\cal N}
\,\exp -S_{+}[x(t)]\,
\exp -\beta W[x(t)].
\end{equation}
The first integral defines the distribution of initial
conditions $x_0$, the integral
$\int {\cal D}_0\,{\cal N}e^{-S_{+}}\cdots$
is over all trajectories launched from a given point
$x_{0}$, each weighted by its probability (Eq.\ref{eq:pab}),
and $e^{-\beta W}$ is the quantity being averaged.
Now, from the definitions of $S_{\pm}$
and $W$ we have:
\begin{eqnarray}
S_+ - S_- + \beta W &=& 
\beta\int_0^\tau dt\,
(\dot x\partial_x V_\lambda + 
\dot\lambda\partial_\lambda V_\lambda) \nonumber\\
&=& \beta\int_0^\tau dt\,
{d\over dt} V_\lambda\Bigl(x(t)\Bigr) \nonumber\\
&=& \beta\Delta V,
\end{eqnarray}
where
\begin{equation}
\Delta V \equiv V_B(x_\tau) - V_A(x_0).
\end{equation}
With Eq.\ref{eq:equivints}, this allows us to
rewrite Eq.\ref{eq:explicit} as
\begin{equation}
\label{eq:reverse}
\overline{\exp -\beta W} =
{1\over Q_{A}}
\int dx_\tau\,\exp -\beta V_{B}(x_\tau)
\int {\cal D}_\tau[x(t)]\,{\cal N}
\exp -S_{-}[x(t)].
\end{equation}
The second integral on the right is unity:
$\int {\cal D}_\tau{\cal N}e^{-S_-} = 1$.
(While this may be verified explicitly,
it also offers a nice interpretation:
if we start in a configuration $x_\tau$
and switch $\lambda$ {\it from $B$ to $A$},
then ${\cal N}e^{-S_-}$ 
is the normalized probability of observing the 
``reverse'' of the trajectory $x(t)$, i.e.\ a 
trajectory which starts at $x_\tau$ and
``evolves backwards'', ending at $x_0$.)
Then, using $Q_{B}=\int dx\,e^{-\beta V_{B}(x)}$,
we finally get
\begin{equation}
\overline{\exp -\beta W} = {Q_{B}\over Q_{A}}
= \exp -\beta\Delta F.
\end{equation}
{\it Q.E.D.}

I have just sketched a proof of Eq.\ref{eq:iden}
for the special case in which the evolution of
the system of interest is modelled by a Langevin
equation in configuration space.
In this proof, an expression for the work $W$
performed on the system, Eq.\ref{eq:work}, was
introduced without elaboration.
More generally, the work is given by
\begin{equation}
\label{eq:genwork}
W = \int_0^\tau dt\,\dot\lambda\,
\partial_\lambda H_\lambda,
\end{equation}
where $H_\lambda$ is the parameter-dependent
Hamiltonian for the system of interest, and
$\partial_\lambda H_\lambda$ in the integrand
is evaluated along a trajectory ${\bf z}(t)$
describing the evolution of the phase space
coordinates of the system.
For Hamiltonians of the form
$H_\lambda=p^2/2m+V_\lambda(x)$, Eq.\ref{eq:genwork}
reduces to Eq.\ref{eq:work}.
Let me now make a few comments regarding the
origin of Eq.\ref{eq:genwork}.

The external work performed on an {\it isolated}
system is equal to the net change in its energy.
When that work is performed by the variation
of an external parameter, we get
\begin{eqnarray}
W &\equiv& H_B({\bf z}_\tau) - H_A({\bf z}_0) 
\nonumber\\ 
&=& \int_0^\tau dt\,{d\over dt}
H_\lambda\Bigl({\bf z}(t)\Bigr) \nonumber\\
\label{eq:hamwork}
&=& \int_0^\tau dt\,\dot\lambda\,
\partial_\lambda H_\lambda\Bigl({\bf z}(t)\Bigr),
\end{eqnarray}
assuming a phase space trajectory ${\bf z}(t)$ 
evolving under Hamilton's equations, so that
$dH/dt=\partial H/\partial t$\cite{goldstein}.
Eq.\ref{eq:genwork} is therefore the correct
expression for work, provided the system is 
isolated during the switching process.

When the system of interest is coupled to a 
heat reservoir, then we may treat the two
together as a larger, isolated system governed
by a Hamiltonian of the form
\begin{equation}
\label{eq:fullham}
{\cal H}_\lambda({\bf z},{\bf z}^\prime) = 
H_\lambda({\bf z}) + H_{res}({\bf z}^\prime)
+ h_{int}({\bf z},{\bf z}^\prime).
\end{equation}
Here, ${\bf z}^\prime$ represents a point in
the phase space of the reservoir, $H_{res}$ is
a Hamiltonian for the reservoir alone,
and $h_{int}$ is a term which weakly couples 
the system of interest to the reservoir.
Since the system of interest and reservoir
together
constitute a larger, isolated system, we
may again use Eq.\ref{eq:hamwork} for the
external work performed, but with
$H_\lambda$ replaced by ${\cal H}_\lambda$.
However, 
$\partial_\lambda{\cal H}_\lambda({\bf z},{\bf z}^\prime)
=\partial_\lambda H_\lambda({\bf z})$,
so we again end up with Eq.\ref{eq:genwork}.
\footnote{In writing Eq.\ref{eq:fullham}, I 
was careful to make only the first term on the
right depend on $\lambda$.
Otherwise, by externally changing $\lambda$, 
we would perform work directly on the degrees 
of freedom of the reservoir, a situation 
different from that considered in this talk.}
Note that the work $W$ no longer represents
the net change in the energy of the system
of interest itself, 
$W\ne H_B({\bf z}_\tau) - H_A({\bf z}_0)$, 
but rather the net change
in the total energy of system and reservoir.

Finally, it is good to keep in mind that
the concept of an external parameter is itself
an idealization.
In reality, such a ``parameter'' must actually
be a degree of freedom --- subject to back 
reaction forces --- and the work performed by
it, over some length of time, is simply the net
{\it loss} in its energy.
The idealization lies in assuming an infinite
inertia for this degree of freedom.
To illustrate these points, consider a
Hamiltonian 
\begin{equation}
G(\lambda,P_\lambda,{\bf z},{\bf z}^\prime)
= {P_\lambda^2\over 2M} + 
{\cal H}_\lambda({\bf z},{\bf z}^\prime),
\end{equation}
where $\lambda$ now represents the degree
of freedom which was previously viewed as
an external parameter, and $M$ and $P_\lambda$
denote the associated inertia and momentum,
respectively.
The first term on the right represents the 
energy of the parametric degree of freedom;
the second, as before, is the Hamiltonian
for the coupled system of interest and 
reservoir.
Hamilton's equations give
\begin{eqnarray}
\label{eq:ham1}
\dot\lambda &=& {\partial G\over\partial P_\lambda} 
= {P_\lambda\over M} \\
\label{eq:ham2}
\dot P_\lambda &=& -{\partial G\over\partial\lambda} 
= -\partial_\lambda{\cal H}_\lambda
({\bf z},{\bf z}^\prime)
= -\partial_\lambda H_\lambda({\bf z}).
\end{eqnarray}
From these we obtain the following expression
for the rate of change of the energy of the
``parameter'':
\begin{equation}
{d\over dt} \,{P_\lambda^2\over 2M} = 
-\dot\lambda\,\partial_\lambda H_\lambda({\bf z}),
\end{equation}
from which, taking $W$ to be {\it minus} the
change in $P_\lambda^2/2M$,
we once again get 
\begin{equation}
W = \int_0^\tau dt\,\dot\lambda\,
\partial_\lambda H_\lambda\Bigl({\bf z}(t)\Bigr).
\end{equation}
Note that now the time-dependence of 
$\lambda$ is not exactly that given by
Eq.\ref{eq:loft}, but rather is determined
from Hamilton's equations.
If, however, we consider the initial conditions
$\lambda(0)=\lambda_A$ and 
$\dot\lambda(0)=(\lambda_B-\lambda_A)/\tau$,
and we take the limit $M\rightarrow\infty$,
then over any finite time interval $\tau$ we will 
get (see Eqs.\ref{eq:ham1} and \ref{eq:ham2})
\begin{equation}
\ddot\lambda = 
{d\over dt}\,{P_\lambda\over M}\rightarrow 0,
\end{equation}
thus recovering Eq.\ref{eq:loft} for the
evolution of $\lambda$.

In the preceding discussion, I have argued that
Eq.\ref{eq:genwork} (which in turn implies
Eq.\ref{eq:work} when $H_\lambda=p^2/2m+V_\lambda$)
is the correct expression for the work performed
on a system by the variation of an external
parameter,
both when the system is isolated and when it is
coupled to a heat reservoir, and also when 
the external parameter is treated honestly
as a degree of freedom (but in the limit of
infinite inertia).
This discussion can be illustrated by considering 
the example of a closed container
filled with gas, where one wall of the container
is free to move in and out as a piston.
Let $\lambda$ denote the position of the piston, 
and ${\bf z}=({\bf x}_1,{\bf p}_1,\cdots,
{\bf x}_N,{\bf p}_N)$ the positions and momenta
of the $N$ individual gas molecules.

When we externally move the piston at some finite
rate from a position $\lambda=A$ to another
position $\lambda=B$, we
perform a quantity of work each time a
gas molecule scatters off the moving wall;
that work is just the change in the kinetic 
energy of the molecule during the collision.
If the container is isolated, then
this is the only mechanism by which the 
energy of the gas can change, so at the end of 
the switching process the work performed is equal 
to the change in the internal energy of the gas.
It is a straightforward exercise\cite{bill} to show 
explicitly (without invoking the Hamiltonian 
identity $dH/dt = \partial H/\partial t$) 
that the change in the energy of the gas,
during a given collision between a molecule and
the moving wall, is equal to the time integral 
of $\dot\lambda \partial_\lambda H_\lambda$
along the trajectory ${\bf z}(t)$ describing
the phase space evolution of the gas, from 
a time immediately before to a time immediately
after the collision.
The total change in energy (and therefore the
work performed) over some finite period
of time is then just the integral of
$\dot\lambda \partial_\lambda H_\lambda$
along ${\bf z}(t)$, over that span of time.

Now imagine that the wall of the container
opposite to the piston is externally maintained
at some temperature $T$.
Then there exist two mechanisms by which 
the energy of the gas changes:
by the scattering of a molecule off the 
moving piston, as above, 
and by the scattering of a molecule off the
thermostatted wall.
As before, the net contribution of the
former is the time integral of
$\dot\lambda \partial_\lambda H_\lambda$
(even though the trajectory ${\bf z}(t)$ is
no longer Hamiltonian).
It is only this contribution which counts
as {\it work} performed on the gas:
the sum of all the energy changes 
due to collisions
with the thermostatted wall is the {\it heat} 
absorbed or relinquished by the gas.
Thus we again obtain Eq.\ref{eq:genwork}.

Finally, suppose the piston is itself a
massive object, moving frictionlessly from
$A$ to $B$, rather than an externally 
pushed device.
As before, work is performed by the
piston every time a molecule scatters off it
(and again the total work is given by the integral 
of $\dot\lambda \partial_\lambda H_\lambda$),
but now the kinetic energy of the piston changes
at each such collision.
If the piston begins with a speed $\dot\lambda_0$,
and is then observed for a time $\tau$,
then the final speed will be
\begin{equation}
\dot\lambda_\tau = 
\sqrt{\dot\lambda_0^2 - 2W/M},
\end{equation}
$M$ being the mass of the piston.
For any set of initial conditions of the gas, we 
get $\dot\lambda_\tau\rightarrow\dot\lambda_0$
in the limit $M\rightarrow\infty$, so in that
limit the speed of the piston remains constant,
just as if it were being driven externally.
(In other words, for an infinitely massive
piston, the work performed by it on the gas
represents an infinitesimal proportional change
in the piston's kinetic energy.)

\section*{CONCLUSIONS}

The focus of this talk has been a 
result (Eq.\ref{eq:iden}) which may be viewed
as an extension --- to irreversible, 
nonequilibrium processes --- of the well-known
relationship between {\it reversible} work
and free energy (Eq.\ref{eq:qstat}).
Just as the reversible work performed in
parametrically switching a system from $A$ to
$B$ is equal to $\Delta F$ regardless of the
path taken in parameter space,
so the average 
$\overline{\exp -\beta W}$ --- defined with
respect to a statistical ensemble of
{\it irreversible} processes --- is equal
to $\exp -\beta\Delta F$ regardless of both
the path taken, {\it and the rate at which the 
switching is carried out}.
After presenting this result, I have
sketched a proof for the special case when
the system evolves under a Langevin equation
in configuration space.
I have also discussed the
general expression for the work performed
on the system (Eq.\ref{eq:genwork}), in
terms of a phase space trajectory ${\bf z}(t)$
describing the evolution of the micro-state
of the system.

\section*{ACKNOWLEDGMENTS}

This research was partially supported by the 
Polish-American Maria Sk\l odowska-Curie Joint 
Fund II, under project PAA/NSF-96-253.
The discussion of work in the final portion of
this paper was stimulated by questions raised
during the presentation of these results at the
Marian Smoluchowski Symposium on Statistical Physics
in Zakopane, Poland, September 1-10, 1997.

\begin{references}

\bibitem{ll}
L.D.Landau and E.M.Lifshitz, {\it Statistical Physics},
3rd ed., Part 1, section 15
(Pergamon Press, Oxford, 1990).

\bibitem{iden}
C.Jarzynski, Phys.Rev.Lett.{\bf 78}, 2690 (1997).

\bibitem{mast}
C.Jarzynski, Phys.Rev.E {\bf 56}, 5018 (1997).

\bibitem{gavin}
G.E.Crooks,``Nonequilibrium measurements of free
energy differences for microscopically reversible 
Markovian systems'', to appear in J.Stat.Phys.

\bibitem{polymers} F.W.Wiegel, {\it Introduction
to Path-Integral Methods in Physics and Polymer
Science} (World Scientific, Philadelphia, 1986).
Original references are:
S.Chandrasekhar, Rev.Mod.Phys.{\bf 15}, 1 (1943);
L.Onsager and S.Machlup, Phys.Rev.{\bf 91}, 1505
(1953) and Phys.Rev.{\bf 91}, 1512 (1953).

\bibitem{goldstein} H.Goldstein, {\it Classical
Mechanics}, 2nd ed., chapter 8.
Addison-Wesley, Reading, Massachusetts, 1980.

\bibitem{bill} 
See, for instance, 
C.Jarzynski, Phys.Rev.E {\bf 48}, 4340 (1993),
Section V.

\end{references}
\end{document}